\documentclass[prb,twocolumn]{revtex4}
\usepackage{amssymb}
\usepackage{graphicx}
\usepackage{amsfonts}
\usepackage{amsmath}

\begin{document}

\title{Scattering theory of magnetic/superconducting junctions with spin active interfaces}

\author{F. Romeo$^{1,2}$ and R. Citro$^{1,2}$}
\affiliation{$^{1}$Dipartimento di Fisica ''E. R. Caianiello'',
Universit{\`a} degli Studi di Salerno, Via Ponte don Melillo,
I-84084 Fisciano (Sa), Italy\\
$^{2}$Institute CNR-SPIN, Via Ponte don Melillo, I-84084 Fisciano (Sa), Italy}

\begin{abstract}
We formulate a generalized scattering field theory \textit{\`{a} la} B\"{u}ttiker describing particles transport in magnetic/superconducting heterostructures. The proposed formalism, characterized by a four-component spinorial wavefunction of the Bogoliubov de Gennes theory, allows to describe the spin flipping phenomena induced by noncollinear magnetizations in the scattering region. As a specific application of the theory, we analyze the conductance, the magnetoresistance and the generation of spin-torque produced by an applied voltage in a spin-valve system. Quantum size effects and quantum beating patterns both in the conductance and in the spin-torque are carefully described.
\end{abstract}



\maketitle

\section{Introduction}

Nanoscale structures involving Normal (N), Ferromagnet (F) and
Superconductor (S) junctions, the so-called heterostructures, involve interplay of superconducting and ferromagnetic order parameters providing a novel opportunity to study the influence of the spin degree of freedom on transport and thermodynamic properties of such systems.
A paradigmatic example is represented by the normal metal (N)/superconducting (S) bilayer. In this system the sub-gap transmission of an electron propagating from the N-side towards the S-side is forbidden due to the absence of available electronic states within the superconducting gap. Thus, in order to conserve the charge current, a propagating Cooper pair is transmitted in the superconductor, while an hole is reflected in the normal metal. This anomalous reflection, described by the pioneering work of Andreev\cite{andreev_original}, has been successively described by Blonder \textit{et al.}\cite{btk82} in the language of scattering theory (the so-called BTK theory).  Differently from methods based on the transfer Hamiltonian formalism (THF), the BTK scattering approach does not assume weak coupling approximation and allows to study transport in NS heterostructures from the metallic (high transparency of the interface) up to the tunneling limit (low transparency). In the tunneling limit, the agreement between BTK theory and the Green's function approach shows that, in the absence of many-body correlations, a scattering theory is extremely suitable for studying transport in heterostructures avoiding time-consuming methods. Almost ten years after the BTK formulation, a development for F/S interfaces based on a BTK-like theory able to describe the Andreev reflection physics within the scattering approach appeared\cite{deJong_Beenakker95}.
A further improvement of the original BTK formalism has been successively brought out by Anantram \textit{et al.}\cite{anantram96} who reformulated the BTK theory in terms of the scattering field theory\cite{buttiker92} (BSFT) originally conceived by B\"{u}ttiker\cite{buttiker92} for mesoscopic (normal) systems. The formalism provides a clear correspondence between physical observables and the scattering matrix of the system within a second quantization formalism, avoiding the by-hand construction characterizing some parts of the BTK theory\cite{lambert91}. The Anantram and Datta work\cite{anantram96} is based on a two-component spinorial Bogoliubov-de Gennes\cite{degennes_book} theory (as in the original BTK theory) which is a convenient representation in the analysis of spin conserving processes. However, recently, the need to explore the interplay between superconductivity and magnetism stimulated the realization of magnetic superconducting heterostructures. In the simplest case of a ferromagnetic F/S bilayer\cite{cuoco}, the bulk magnetization of the F-side can differ from the one at the interface and spin flipping phenomena can take place at the F/S junction. In order to fully describe the physics of the magnetic superconducting heterostructures one needs to generalize the BdG formalism to a four-component spinorial representation. To the best of our knowledge, a scattering field theory \textit{\`{ a} la} B\"{u}ttiker for magnetic superconducting heterostructures which properly treat spin active phenomena at the interface is not yet available and our work aims to fill such vacancy.
In particular, the second quantized form of the BTK theory offers important advantages in obtaining non-local quantum properties (i.e. correlation functions) which are not provided by the BTK theory.  Following the work of Anantram \textit{et al.}\cite{anantram96}, the construction of a spinful theory is important to correctly describe interplay phenomena between superconductivity and magnetism.

For the presentation we follow the formalism of Refs.[\onlinecite{btk82},\onlinecite{anantram96}] and focus our attention
on quasi-one-dimensional systems, while the generalization to the 2-dimensional case is left for a forthcoming work. Within our generalized scattering
formalism we derive the expression for the charge and spin currents and the respective linear response  to an external bias, i.e. the conductance and the spin-torque. The second part of the paper is devoted to the study of the spin polarized transport in a spin-valve system by the developed formalism. In particular, we focus on the evidence of the Andreev reflection in the subgap transport at varying the magnetic interaction and on the quantum size effects due to the interlayer width to make a link to experimental observations. Among the transport properties, we do analyze the conductance, the magnetoresistance and finally the spin torque as a probe of spin-polarized transport.

The organization of the paper is the following: In Sec.\ref{sec:model} we introduce the Bogoliubov-de Gennes Hamiltonian and present
the scattering field theory generalized to include spin flipping phenomena.
We then derive the expression of the charge and spin current and of the respective differential conductance in the presence of an external bias.
In Sec.\ref{sec:results} we present the results of the linear response observables: the conductance, the magnetoresistance and the torkance for the structure shown in Fig.\ref{fig:fig2}. Conclusions and perspectives are given in Sec.IV.

\section{ The model and formalism}
\label{sec:model}
\begin{figure}[t]
\centering
\includegraphics[scale=0.5]{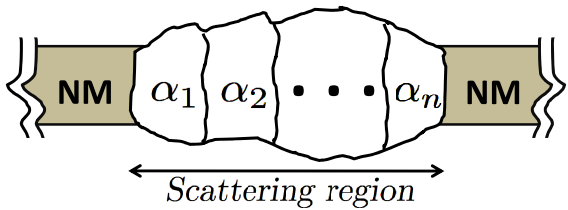}
\caption{Magnetic superconducting heterostructure as described in the main text. The sequence of materials $[\alpha_1|\alpha_2|...|\alpha_n]$ can be $\alpha_n \in\{F, S, NM\}$.}
\label{fig:fig1}
\end{figure}
We consider a one-dimensional magnetic superconducting heterostructure connected to normal nonmagnetic leads (see Fig.\ref{fig:fig1}). The scattering region is made by magnetic and superconducting regions $\alpha_i$  arbitrarily disposed along the transport direction. The system is conveniently described by a 4-component Bogoliubov-de Gennes (BdG) formalism in which the quantum state of the system is described by the wave-function $|\Psi(x,t)\rangle=\sum_{\beta,\sigma}\phi_{\beta \sigma}(x,t)|\beta\rangle\otimes|\sigma\rangle$, where $\beta \in\{e,h\}$ is the particle index and $\sigma \in \{+,-\}$ represents the spin orientation along the quantization axis. Introducing the standard notation of the BdG theory we set $\phi_{e\sigma}(x,t) \rightarrow u_{\sigma}(x,t)$ and $\phi_{h\sigma}(x,t) \rightarrow v_{\sigma}(x,t)$ thus the state vector can be put in the form $|\Psi(x,t)\rangle=(u_{\uparrow}(x,t),u_{\downarrow}(x,t),v_{\uparrow}(x,t),v_{\downarrow}(x,t))^t$.\\

Using a tensor product notation (see APPENDIX \ref{app:tensor-product}), the quasiparticle Hamiltonian can be represented as\cite{nota1}
\begin{eqnarray}
\label{eq:bdg_ham}
\hat{H} &=& P_{ee} \otimes (\hat{H}_e+ \hat{U}(x))+P_{hh} \otimes [-(\hat{H}_e+ \hat{U}(x))^{\ast}]\nonumber \\
&+&P_{eh} \otimes \hat{\Delta}(x)+P_{he} \otimes \hat{\Delta}(x)^{\dagger},
\end{eqnarray}
where $P_{\alpha\beta}=|\alpha\rangle\langle\beta|$ ($\alpha,\beta=e,h$) are particle and/or hole \textit{projectors} with $|e\rangle=(1,0)^t$ and $|h\rangle=(0,1)^t$;  $\hat{H}_e$, $\hat{U}(x)$, $\hat{\Delta}(x)$ are operators written in the spin-space basis $|\sigma=\pm\rangle$, $|+\rangle=(1,0)^t$ and $|-\rangle=(0,1)^t$ ($t$ stands for the transposed vector). Specifically,  $\hat{H}_e$ is kinetic energy operator defined by
\begin{equation}
\hat{H}_e=\Bigl[-\frac{\hbar^{2} \partial^{2}_x}{2m}-E_F \Bigl]\hat{\mathbb{I}}_{sp},
\end{equation}
being $E_F$ the Fermi energy and $\hat{\mathbb{I}}_{sp}$ the identity operator in the spin space. The potential energy $\hat{U}(x)$ may include spin dependent Zeeman or spin-orbit coupling terms (e.g. $\hat{U}(x)=\vec{h}(x)\cdot \hat{\vec{\sigma}}$) and finally for an s-wave superconductor, $\hat{\Delta}(x)=i\hat{\sigma_y}\Delta(x)$ where $\sigma_y$ is a Pauli matrix. In the following we characterize the superconducting regions by a constant order parameter $\Delta$ (i.e.  we assume a step-like behavior of the gap at the N/S interfaces\cite{bozovic}).

Within our tensor product representation, the charge and spin current operator in first quantization can be written in a compact form as:
\begin{equation}
\hat{J}_\mu=\frac{i\hbar q_{\mu}}{2m}\Bigl(\overleftarrow{\partial}_x-\overrightarrow{\partial}_x \Bigl)\sum_{\beta}\eta_\beta P_{\beta\beta}\otimes\sigma_{\mu}^{\beta},
\end{equation}
where the index $\mu=0,1,\ldots,3$, so that $\hat{J}_0$ is the charge current operator, while $\hat{J}_1$,  $\hat{J}_2$ and  $\hat{J}_3$ represent the three space-components of the spin current operator. In the above notation we introduced $q_{0}=q_e$, $q_1=q_2=q_3=\hbar/2$, $\eta_e=-\eta_h=1$ and $(\sigma^h_{\mu})^\ast=\sigma^e_{\mu}=\sigma_{\mu}$, $\sigma^{\beta}_0=\eta_{\beta}\mathbb{I}_{sp}$, being $q_e=-|e|$ the electron charge and $\sigma_{\mu}$ the $\mu$-th Pauli matrix.\\
The above expression for the currents derives from the conservation laws of the charge $\hat{Q}$ and spin $\hat{S}_{\mu}$ densities,
\begin{eqnarray}
\hat{Q}&=&q_e\sum_\beta \eta_{\beta} P_{\beta\beta}\otimes \hat{\mathbb{I}}_{sp}\\\nonumber
\hat{S}_{\mu}&=&(\hbar/2)\sum_\beta  P_{\beta\beta}\otimes \hat{\sigma^{\beta}}_{\mu}.
\end{eqnarray}
In particular, the charge and spin densities conservation law can be put in the form of a continuity equation with source/sink terms. Concerning the charge continuity equation, the source/sink term is related to the divergence of the Cooper pairs current\cite{yamashita03}, while  in the spin density case such term is related to the spin-torque. Indeed, a spin-torque $\hat{T}_{\mu}$ can be generated by a magnetic potential of the form $\hat{U}(x)=\vec{h}(x)\cdot \hat{\vec{\sigma}}$ and its $\mu$-component is represented by the operator
\begin{equation}
\label{eq:torque-locale}
\hat{T}_{\mu}=\sum_{\beta} P_{\beta\beta} \otimes \Bigl[\vec{h}(x)\times \hat{\vec{\sigma}}^{\beta}\Bigl]_{\mu}.
\end{equation}
The continuity equation for the spin density can thus be written as $\partial_t \hat{S}_{\mu}+\partial_x \hat{J_{\mu}}=\hat{T}_{\mu}$.

\subsection{Scattering formalism}
In the following we describe the scattering field theory for magnetic/superconducting heterostructures.
In constructing the theory, the Andreev approximation (which neglects the difference between the particle and hole momentum) is performed only in the external leads, while inside the scattering region the exact wave-functions are considered. This approach is performed to properly treat the phase-coherent phenomena in the scattering region and to correctly capture all the Andreev-reflections probabilities.\\
The scattering field (see APPENDIX \ref{app:scattering-field}) in the $j$-th lead can be written as\cite{nota2}:
\begin{eqnarray}
\label{eq:scattering fields}
\hat{\Psi}_j(x,t)&=&\sum_{\beta,\sigma}\int \frac{dE\exp[-iEt]}{\sqrt{2 \pi \hbar v(E)}}|\beta \rangle \otimes |\sigma \rangle \times\\\nonumber
&[& a^{\sigma}_{j\beta}(E)e^{i k_{\beta}x}+b^{\sigma}_{j\beta}(E)e^{-i k_{\beta}x}],
\end{eqnarray}
where $k_{\beta}=\eta_{\beta}k(E)$, while the scattering operator $a^{\sigma}_{j\beta}(E)$ ($b^{\sigma}_{j\beta}(E)$) destroys an incoming (outgoing) particle of species $\beta \in \{e , h \}$ and spin projection $\sigma \in \{+,-\}$ in the lead $j$
and within the  Andreev approximation $v_{j\beta\sigma}(E) \approx v(E)=\hbar k(E)/m$. The scattering field defined above generalizes the one introduced in Ref.[\onlinecite{anantram96}]  to the spinful case.
The outgoing field operators are related to the incoming field operators by the scattering matrix\cite{nota3,nam_d02007}:
\begin{equation}
b^{\sigma}_{i\beta}(t)=\sum_{i'\sigma'\beta'}S^{\beta\beta'}_{ii'\sigma\sigma'} (t) a^{\sigma'}_{i'\beta'}(t).
\end{equation}
Since the scattering states given by (\ref{eq:scattering fields}) form a complete set of mutually orthogonal states (completeness relation),
the fields $b_i^\sigma$ satisfy the canonical commutation relations $\{b^{\sigma}_{i\beta},(b^{\sigma'}_{i'\beta'})^{\dagger}\}=\delta_{ii'}\delta_{\beta\beta'}\delta_{\sigma\sigma'}$ and the current conservation ensures the unitary condition of  the $S$-matrix:
\begin{equation}
\sum_{i'\sigma'\beta'}S^{\beta\beta'}_{ii'\sigma\sigma'}(S^{b\beta'}_{ki's\sigma'})^{\ast}=\delta_{ik}\delta_{b\beta}\delta_{s\sigma}.
\end{equation}
The quantum statistical properties of the leads are defined by the expectation value $\langle a^{\sigma\dagger}_{j\alpha}(E) a^{s}_{i\beta}(E') \rangle=\delta_{ij}\delta_{s\sigma}\delta_{\alpha\beta}\delta(E-E')f_{j\alpha}(E)$, $f_{j\alpha}(E)$ being the Fermi distribution of the particle of species $\alpha$ in the electrode $j$. \\
The field operator $\hat{\Psi}_j$ acts on a many-particle state and thus the expectation value $\bar{\mathcal{O}}_j$ of the generic operator $\hat{\mathcal{O}}$ in the $j$-th electrode is given by $\bar{\mathcal{O}}_j=\langle \hat{\Psi}^{\dagger}_j\hat{\mathcal{O}}\hat{\Psi}_j\rangle$, where the notation $\langle \cdot\cdot\cdot\rangle$ stands for the quantum statistical average.

\subsection{Charge current and differential conductance}
\label{sec:charge current & differential conductance}
In this Section we derive the two-terminal conductance of the magnetic/superconducting heterostructure depicted in Fig.\ref{fig:fig1}. For a multi-terminal device, the average charge current $\bar{J}^{i}_0$ flowing through the $i$-th lead is given by  $\bar{J}^{i}_0=\langle \hat{\Psi}^{\dagger}_i(x,t) \hat{J}_0\hat{\Psi}_i(x,t)\rangle$ and using the field representation (\ref{eq:scattering fields}), it can be expressed in terms of the scattering matrix as
\begin{equation}
\label{eq:charge current}
\bar{J}^{i}_0= \frac{q_e}{h}\sum_{\beta \alpha j}\eta_{\beta}\int dE \Bigl[2\delta_{ij}\delta_{\alpha \beta}-\mathcal{M}_{ij}^{\beta \alpha}(E)\Bigl]f_{j\alpha}(E),
\end{equation}
where $\mathcal{M}_{ij}^{\beta \alpha}(E)=Tr[S^{\beta \alpha \dagger}_{ij}(E)S^{\beta \alpha }_{ij}(E)]$, $Tr[\cdot \cdot \cdot]$ indicates the trace on the spin indices, while $S^{\beta \alpha }_{ij}(E)$ are matrices with respect to the spin indices.
When a symmetric potential drop is applied to the system, the electrochemical potential in the $i$-th lead can be written as $\mu_i=\mu+(-)^i q_eV/2$, being $\mu=(\mu_1+\mu_2)/2$ and $V$ the bias voltage. Taking as zero of the energies the electrochemical potential of the scattering region $\mu_s$, we can write $f_{j\alpha}(E)=f([E+\eta_{\alpha}(\mu_s-\mu_j)]/(K_B T))$, being $T$ the temperature. Let us note that  $\mu_s=(\mu_1+\mu_2)/2$ only in the symmetric case. In the nonsymmetric case
$\mu_s\neq (\mu_1+\mu_2)/2=\mu$ and thus $\mu_s$ must be determined self-consistently to conserve the charge current (i.e. $\sum_i\bar{J}^{i}_0(V,\mu_s(V))=0$) as described in the Appendix \ref{app:conductance-tensor}.
Within the linear response theory, the charge current flowing through the $i$-th lead is obtained as $I_i=\sum_j g_{ij} (\mu_j-\mu_s)=G_i V$ where $g_{ij}$ is the conductance tensor whose expression (see Appendix \ref{app:conductance-tensor}) is:
\begin{eqnarray}
g_{ik}&=& \frac{e^2}{h}\int d\xi [-\partial_{\xi}f(\xi)]_{eq}\times \\\nonumber
&[& 4\delta_{ik}+\mathcal{M}^{he}_{ik}(\xi)+\mathcal{M}^{eh}_{ik}(\xi)-\mathcal{M}^{ee}_{ik}(\xi)-\mathcal{M}^{hh}_{ik}(\xi)],
\end{eqnarray}
where the sum rule $\sum_{j\alpha}\mathcal{M}^{\beta\alpha}_{ij}(E)=2$ has been used.
For a generic structure the two-terminal conductance in terms of the conductance tensor is given by:
\begin{equation}
\label{eq:two-probe cond}
G=\frac{g_{22}g_{11}-g_{21}g_{12}}{\sum_{ij}g_{ij}}.
\end{equation}
Let us note that in the symmetric case the above relation can be simplified as $G_{sym}=(g_{11}-g_{12})/2$ and thus:
\begin{eqnarray}
\label{eq:g_sym}
G_{sym}&=& \frac{e^2}{h}\int d\xi [-\partial_{\xi}f(\xi)]_{eq}\times \\\nonumber
&[& \mathcal{M}^{ee}_{12}(\xi)+\mathcal{M}^{hh}_{12}(\xi)+\mathcal{M}^{he}_{11}(\xi)+\mathcal{M}^{eh}_{11}(\xi)].
\end{eqnarray}

\subsection{Spin currents and spin-torque}
\label{sec:spin currents & spin-torque}

When the heterostructure contains magnetic layers apart the superconducting ones, the application of an external bias produces a spin current coexisting with  the charge flux. Differently from the charge, the spin density is not conserved and thus a spin torque is exerted along the nanostructure.
In particular the spin torque results from the divergence of the spin current  as discussed in Ref.[\onlinecite{nostri_torque}].

In order to derive an expression of the spin-torque, we need first to derive an expression of the spin current in terms of the scattering matrix.
This is given by the quantum average of the spin current density operator $\bar{J}^i_{\mu}=\langle \hat{\Psi}_i^\dagger \hat{J}_{\mu}\hat{\Psi}_i \rangle$ ($\mu=1,2,3$) and using (6) one obtains:
\begin{equation}
\label{eq:spin-curr}
\bar{J}^i_{\mu}=-\sum_{\alpha \beta j}\int \frac{d E}{4\pi}Tr[S^{\beta \alpha \dagger}_{ij}(E)\sigma^{\beta}_{\mu}S^{\beta \alpha}_{ij}(E)]f_{j\alpha}(E),
\end{equation}
where $\bar{J}^i_{\mu}$ represents the $\mu$-component of the spin current density generated in the $i$-th lead. Let us note that Eq.(\ref{eq:spin-curr}) represents an expression of the spin current beyond the linear response regime.
Moreover in calculating the spin current, the charge current conservation through the system must be monitored.
 In fact the spin current is not conserved due to the presence of a spin-transfer torque acting on the local magnetic momentum of the magnetic region and thus a  violation of the charge conservation law may artificially change the spin current gradient.\\

In order to relate the spin current gradient to the spin torque we can consider the continuity equation of the spin density $S_{\mu}$:
\begin{equation}
\label{eq:cont_torque}
\partial_t S_{\mu}+\vec{\nabla} \cdot \vec{J}_{\mu}=T_{\mu}.
\end{equation}
Under stationary condition, i.e. $\partial_t S_{\mu}=0$, one can apply the Gauss-Green theorem to Eq.(\ref{eq:cont_torque}).
 Let us consider a cylindrical surface of volume $\Omega$ encircling the scattering center and with axis collinear to the transport direction $\hat{x}$:
\begin{equation}
\int_{\Omega} \vec{\nabla} \cdot \vec{J}_{\mu}dV=\int_{\Omega} T_{\mu}dV=\oint_{\Sigma}J_{\mu}\hat{x} \cdot d\vec{s}.
\end{equation}
where $\Sigma$ is the cylindrical surface $\Sigma=s_{1} \cup s_{2} \cup s_{l}$ with $s_{1}$ and $s_{2}$ the areas collinear to the transport direction ($\hat{s}_1=-\hat{s}_2=\hat{x}$) and $s_l$ the lateral surface of the cylinder. Since for a quasi-one dimensional system, the physical quantities  $J_{\mu}$ and $T_{\mu}$ can be considered uniform along the radial direction, one obtains:
\begin{equation}
\label{eq:kir-spin-curr}
\sum_i J^{i}_{\mu}+\int_{SR} T_{\mu}dx=0,
\end{equation}
where the integral is performed over the scattering region (SR). Eq.(\ref{eq:kir-spin-curr}) is the \textit{Kirchhoff's law} for the spin current and it can be used together with Eq.(\ref{eq:spin-curr}) to derive the total spin torque $\tau_{\mu}=\int_{SR} T_{\mu}dx$ produced along the system:
\begin{equation}
\tau_{\mu}=\sum_{\alpha \beta j i}\int \frac{d E}{4\pi}Tr[S^{\beta \alpha \dagger}_{ij}(E)\sigma^{\beta}_{\mu}S^{\beta \alpha}_{ij}(E)]f_{j\alpha}(E).
\end{equation}
However, in spin valve devices it is important to resolve the spatial dependence of the spin torque density. Indeed, a spin valve is a device made by two magnetic layers separated by an interstitial nonmagnetic region (spacer). The relative volume of the magnetic layers can be taken very different: one is the pinned magnetic layer (fixed layer), having the largest volume, and the other is the thin magnetic layer called free-layer (FL).
When a spin-polarized current interacts with the thin ferromagnetic layer it undergoes a spin-filtering and the result is, in general, that a spin-transfer torque is applied to the magnetic layer.
The energy required to change the magnetization direction of the free-layer can be provided by the flux of spin polarized current activated by the application of an external driving field (e.g. dc voltage bias or ac modulations).

In order make an explicit calculation we consider a spin valve, as the one of Fig.2, with a Zeeman potential of the form:
\begin{equation}
\label{eq:pot-mag}
U(x)=[\gamma \delta(x)\hat{n}_1+h(x)\hat{n}_2 ]\cdot \vec{\sigma},
\end{equation}
where $\hat{n}_1=(\sin(\theta),0,\cos(\theta))$ represents the magnetization direction of the FL, $\hat{n}_2=(0,0,1)$ is the direction of the magnetization of the fixed layer, while $h(x)$ is a step-like function. In this model the  free-layer is directly connected to the first lead (i.e. the region $x<0$) and thus the spin torque acting on this region is simply related to the non-equilibrium spin density produced by the bias in the first lead. Using Eq.(\ref{eq:torque-locale}), the zero temperature spin torque components acting on the free-layer in the linear response regime can be written as follows:
\begin{eqnarray}
\label{eq:torque-free-layer}
&& T_{\parallel}=-\frac{eV \Gamma}{4\pi}\sum_{\alpha \beta j}Tr[S^{\beta\alpha\dagger}_{1j} \sigma^{\beta}_y S^{\beta\alpha}_{1j}]\eta_{\alpha}\lambda_{j}\\\nonumber
&& T_{\perp}=\frac{eV \Gamma}{4\pi}\sum_{\alpha \beta j}Tr[S^{\beta\alpha\dagger}_{1j}\Bigl(\sin(\theta) \sigma^{\beta}_z-\cos(\theta) \sigma^{\beta}_x \Bigl)S^{\beta\alpha}_{1j}]\eta_{\alpha}\lambda_{j},
\end{eqnarray}
where $\Gamma=(k_F \gamma)/E_F$, while the coefficients $\lambda_j$ are given in APPENDIX \ref{app:conductance-tensor}. Furthermore,  $T_{\parallel}$ and $T_{\perp}$ are the components of the spin torque parallel and perpendicular to the plane of the magnetization of the fixed layer (i.e. $\vec{T}=T_{\parallel} \hat{\nu}_{\parallel}+T_{\perp} \hat{\nu}_{\perp}$) whose directions are defined by the vectors\cite{nostri_torque}:
\begin{eqnarray}
&& \hat{\nu}_{\parallel}=-\hat{x}\cos(\theta)+\hat{z}\sin(\theta)\\\nonumber
&& \hat{\nu}_{\perp}=\hat{y}.
\end{eqnarray}

\begin{figure}[t]
\centering
\includegraphics[scale=0.45]{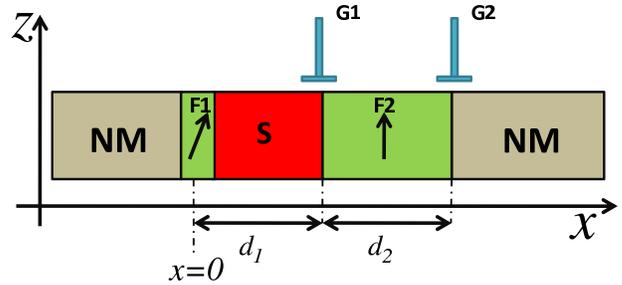}
\caption{Superconducting spin valve as described in the main text.}
\label{fig:fig2}
\end{figure}

\section{Spin-polarized transport in spin-valve systems}
\label{sec:results}
The scattering formalism  developed so far can be employed to describe the linear response properties
(i.e. conductance and torkance) of the superconducting spin valve depicted in Fig.\ref{fig:fig2} (see also APPENDIX \ref{app:boundary}).
The system is described by the BdG Hamiltonian given in Eq.(\ref{eq:bdg_ham}), where the superconducting gap operator is
$\hat{\Delta}(x)=i\sigma_y \Delta \theta(x)\theta(d_1-x)$, $\theta(x)$ being the Heaviside step function while the two magnetic regions,
namely F1 and F2, are modeled by the Zeeman potential given in Eq.(\ref{eq:pot-mag}), where $h(x)=E_F h_z \theta(x-d_1)\theta(d_1+d_2-x)$.
Furthermore an additional barrier potential of the form $\hat{U}_s(x)=\sum_{j=0,1,2}V_j \delta(x-x_j)\mathbb{I}_{sp}$ ($x_0=0$, $x_1=d_1$, $x_2=d_1+d_2$)
is introduced at the interfaces. The $V_j$ are related to the dimensionless BTK parameters $z_j=2mV_j/(\hbar^2 k_F)$ which measure the interface transparencies, and are given by the transmission and reflection probabilities  $\mathcal{T}_j$, $\mathcal{R}_j$
via the relation $z_j=\sqrt{\mathcal{R}_j/\mathcal{T}_j}$. Finally, the s-wave order parameter is taken in dimensionless form as $\eta=\Delta/E_F$ and its relation to the BCS coherence length $\xi$ is given by $k_F \xi=1/\eta$, $\xi \approx \hbar v_F/(2\Delta)$.
In the following we set $\eta=1/200$ and $k_F \approx 1 $\AA$^{-1}$ which
are suitable phenomenological values for conventional superconducting materials such as Nb\cite{yamashita03}.

In the subsequent analysis, adopting the same line of Ref.[\onlinecite{anantram96}], the self-consistent computation of the superconducting order parameter is neglected. In fact we will consider the low-bias regime (i.e. $eV/\Delta \ll 1$) under which the spin accumulation in the superconducting region is unable to produce a relevant suppression of the superconducting gap.

A different mechanism of modification of the superconducting gap could be induced by the size of the superconducting region as reported in Ref.[\onlinecite{bozovic}]. However, as shown in Fig.2 of that work, the superconducting gap saturates to the bulk value as a function of the thickness of the superconducting layer already at values of $2 \xi$ ($\xi$  being the BCS coherence length). These features are quite generic and seem to be robust for any value of the scattering potential at the F/S interface and also for parallel or anti-parallel magnetizations in the ferromagnetic leads. Thus we conclude that in our analysis neglecting the self-consistency of the gap does not induce quantitative important changes.

\subsection{Differential conductance and magnetoresistance}

\begin{figure}[t]
\centering
\includegraphics[scale=0.85]{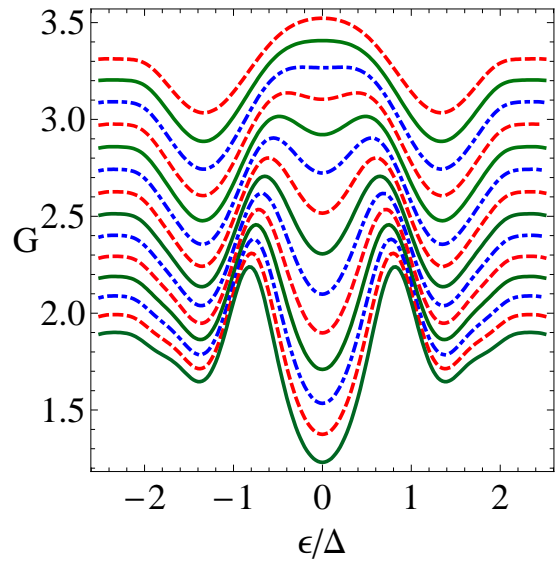}
\caption{Two-probe differential conductance $G$ (in unit of $e^2/h$) as a function of $\epsilon/\Delta$ for the model parameters: $k_F d_1=600$, $k_F d_2=250$, $z_1=z_2=z_3=0.1$, $\theta=0$, $\eta=1/200$, $h_z=-0.45$. The parameter $\Gamma$ takes values ranging from $0.9$ (top curve) up to $2.2$ (bottom curve) and it is increased with constant step of $0.1$ going from top to bottom curve.}
\label{fig:fig3}
\end{figure}

\begin{figure}[h]
\centering
\includegraphics[scale=0.85]{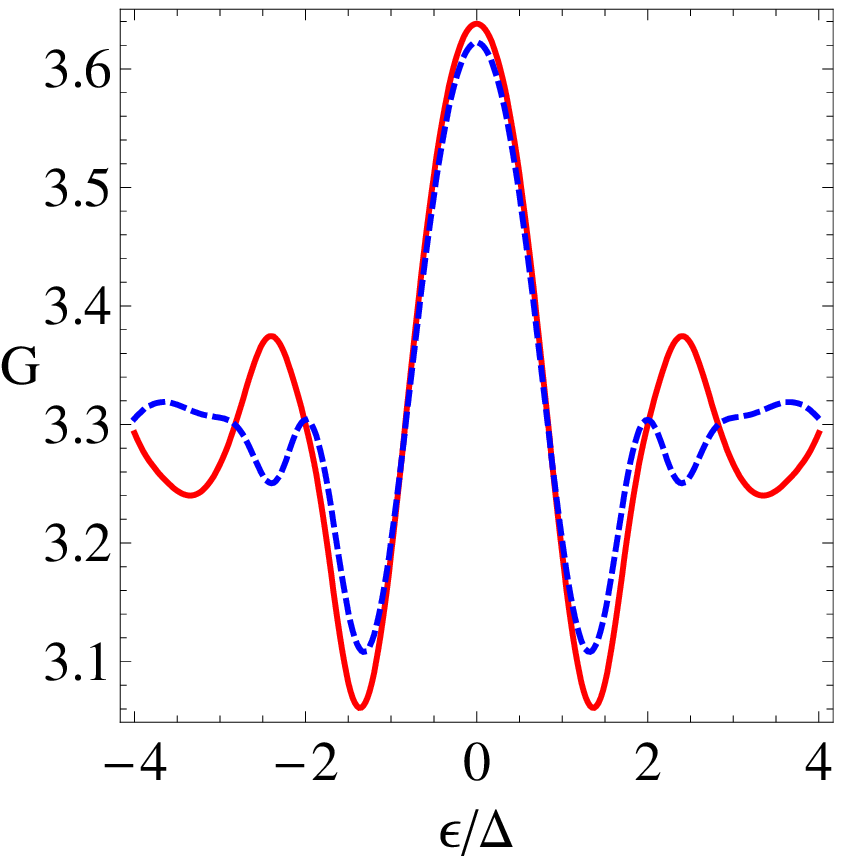}\\
\includegraphics[scale=0.85]{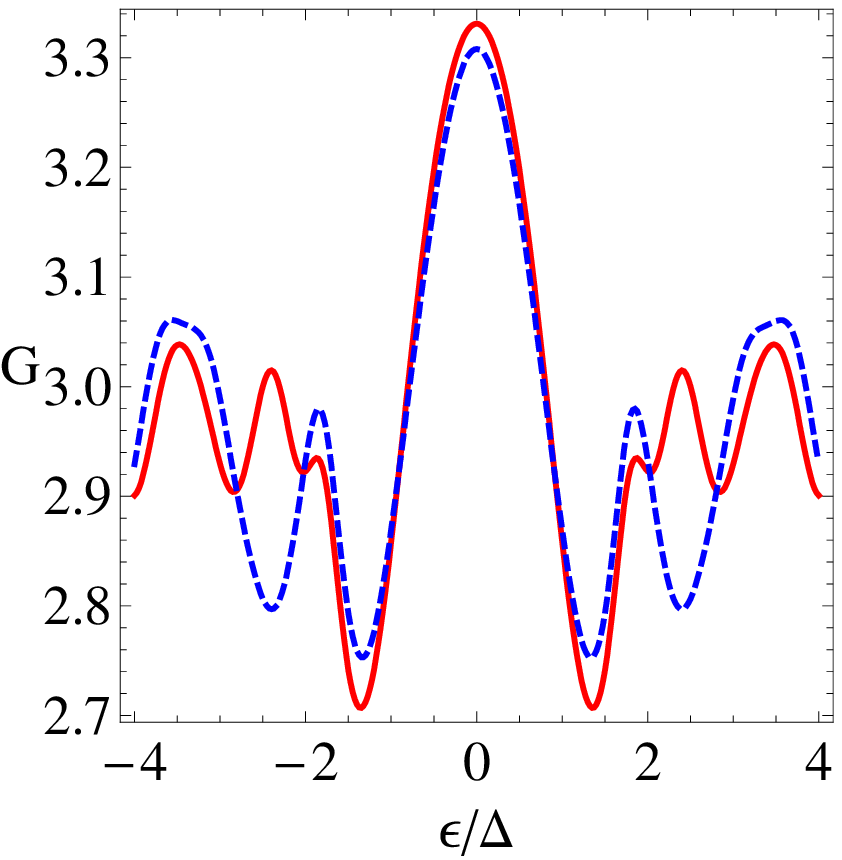}
\caption{Two-probe differential conductance $G$ (in unit of $e^2/h$) as a function of $\epsilon/\Delta$ for the model parameters: $k_F d_1=575$, $k_F d_2=350$, $z_1=z_2=z_3=0.1$, $\Gamma=0.85$, $\eta=1/200$. The full line is computed by setting $\theta=\pi/2$, while the dashed line is obtained fixing $\theta=0$. The Zeeman energy term of the fixed layer is taken $h_z=-0.5$ in the upper panel and $h_z=-0.75$ in the lower one.}
\label{fig:fig4ab}
\end{figure}

\begin{figure}[h]
\centering
\includegraphics[scale=0.55]{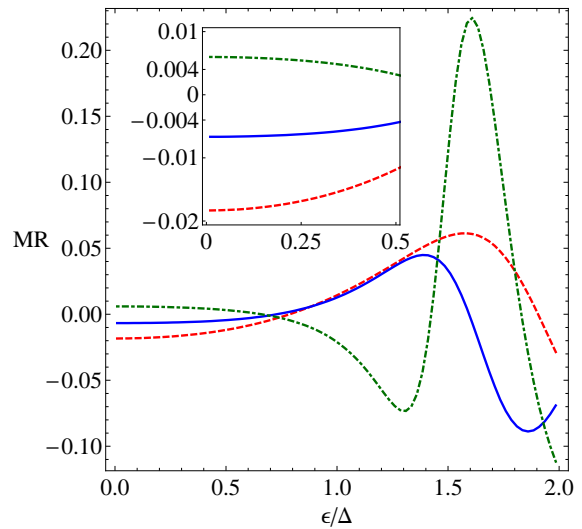}
\caption{Magnetoresistance MR as a function of $\epsilon/\Delta$ for the model parameters as follows: $k_F d_2=250$, $z_1=z_2=z_3=0$, $\Gamma=0.5$, $\eta=1/200$, $\theta=0$, $|h_z|=0.5$, while $k_F d_1=600$ (dashed line),  $k_F d_1=800$ (full line) or $k_F d_1=1000$ (dashed-dotted line). The inset contains the MR behavior within the energy range $[0,0.5]$.}
\label{fig:fig5}
\end{figure}

\begin{figure}[h]
\centering
\includegraphics[scale=0.55]{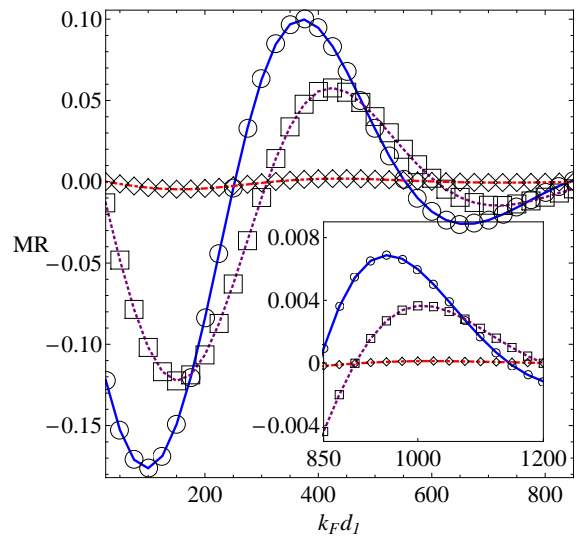}\\
\caption{Magnetoresistance MR as a function of the size $k_F d_1$ of the superconducting region computed for the model parameters: $\epsilon/\Delta=0.01$, $z_1=z_2=z_3=0$, $\Gamma=0.5$, $\eta=1/200$, $|h_z|=0.5$, $\theta=0$, while $k_F d_2=250$ (circle, $\circ$),  $k_F d_2=300$ (square, $\square$) or $k_F d_2=400$ (diamond, $\lozenge$). The inset contains the MR \textit{vs} $k_Fd_1$ in the range $[850,1200]$. The sampling step is 25 (i.e. $2.5$nm).}
\label{fig:fig6}
\end{figure}

\begin{figure}[h]
\centering
\includegraphics[scale=0.55]{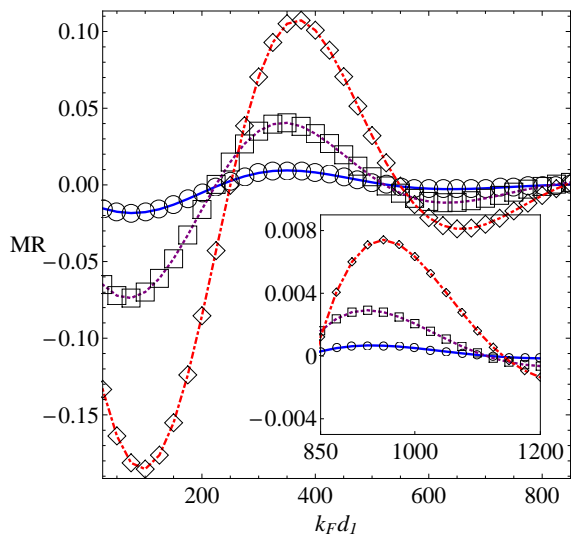}\\
\caption{Magnetoresistance MR as a function of the size $k_F d_1$ of the superconducting region computed for the model parameters: $\epsilon/\Delta=0.01$, $z_1=z_2=z_3=0$, $\Gamma=0.5$, $\eta=1/200$, $\theta=0$, $k_F d_2=650$, while  $|h_z|=0.1$ (circle, $\circ$),  $|h_z|=0.25$ (square, $\square$) or $|h_z|=0.5$ (diamond, $\lozenge$). The inset contains the MR \textit{vs} $k_Fd_1$ in the range $[850,1200]$. The sampling step is 25 (i.e. $2.5$nm).}
\label{fig:fig7}
\end{figure}

\begin{figure}[h]
\centering
\includegraphics[scale=0.6]{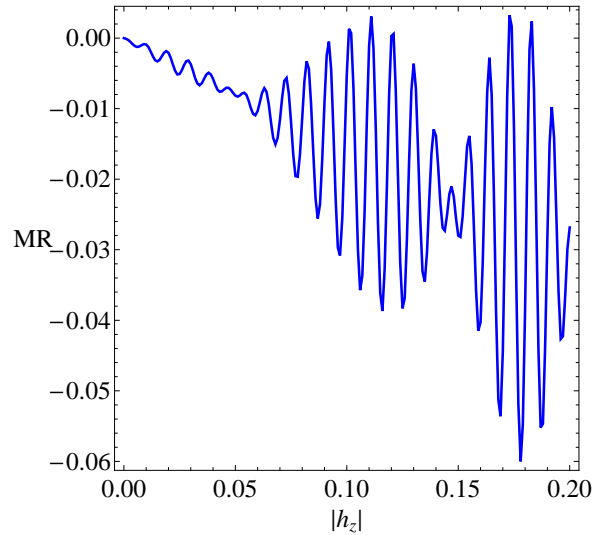}\\
\caption{Magnetoresistance MR as a function of  $|h_z|$ computed for the model parameters: $\epsilon/\Delta=0.01$, $z_1=z_2=z_3=0$, $\Gamma=0.5$, $\eta=1/200$, $\theta=0$, $k_F d_2=650$, $k_F d_1=200$.}
\label{fig:fig8}
\end{figure}

\begin{figure}
\centering
\includegraphics[scale=0.62]{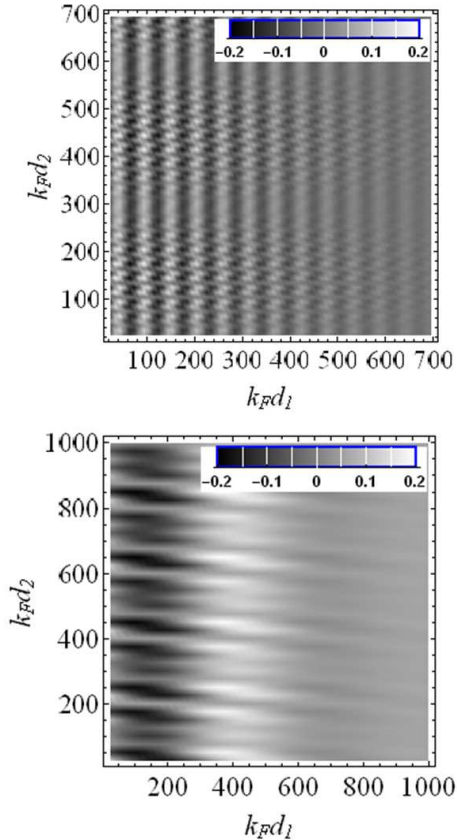}
\caption{Magnetoresistance MR as a function of $k_F d_1$ and $k_F d_2$ for the model parameters: $\epsilon/\Delta=0.01$, $z_1=z_2=z_3=0$, $\Gamma=0.5$, $\eta=1/200$, $\theta=0$ and $|h_z|=0.3$ (upper panel) or $|h_z|=0.5$ (lower panel). The density-plot shows the large scale structure of the oscillations.}
\label{fig:fig9ab}
\end{figure}

In Fig.\ref{fig:fig3} we report the differential conductance $G$ as a function of the energy $\epsilon/\Delta$, with $\epsilon=eV$,
computed setting the model parameters as follows: $k_F d_1=600$, $k_F d_2=250$, $z_1=z_2=z_3=0.1$, $\theta=0$, $\eta=1/200$, $h_z=-0.45$.
At increasing the Zeeman interaction $\Gamma$ of the thin layer, a lowering of the conductance is observed below the gap. This is due to the fact that Andreev reflection processes
dominating the transport properties below the superconducting gap become suppressed. This behavior qualitatively reproduces
 the experimental observations reported in Ref.[\onlinecite{soulen98_AR}] by STM technique. The effect of the spin active barrier
 on the transport properties of the system is analyzed in Figs.\ref{fig:fig4ab}.
The figures represent the differential conductance $G$ computed using the parameters:  $k_F d_1=575$, $k_F d_2=350$, $z_1=z_2=z_3=0.1$, $\Gamma=0.85$, $\eta=1/200$. For both the upper and lower panel the full line is computed by setting $\theta=\pi/2$, while the dashed line is obtained fixing $\theta=0$. The Zeeman energy of the fixed layer is taken $h_z=-0.5$ in the upper panel and $h_z=-0.75$ in the lower one. The analysis of the figures shows that the sub gap transport is not much sensitive to the magnetization direction, while the quasi-particles transport depends on the orientation of the magnetization of the free-layer and more harmonics appear in the oscillating behavior of $G$ above the gap. We do also observe a lowering of the differential conductance as a function of $h_z$  from the upper to the lower panel.
The origin of the oscillations above $\epsilon \simeq \Delta$ is due to releasing the Andreev approximation and are related to the formation of quasiparticles resonances above the gap\cite{dong2003}.\\
In order to describe the magneto-transport properties of the system, we introduce the magnetoresistance (MR) defined as follows:
$MR=[G_{P}-G_{AP}]/G_{AP}$, where we defined $G_{P}=G(|h_z|,\Gamma,\theta=0)$ and $G_{AP}=G(-|h_z|,\Gamma,\theta=0)$.
In Fig.\ref{fig:fig5} we report the MR as a function of $\epsilon/\Delta$ setting the remaining parameters as
follows: $k_F d_2=250$, $z_1=z_2=z_3=0$, $\Gamma=0.5$, $\eta=1/200$, $\theta=0$, $|h_z|=0.5$.
The different curves are related to different width of the superconducting region and in particular the dashed line indicates $k_F d_1=600$,  the full line $k_F d_1=800$, while the dashed-dotted line $k_F d_1=1000$. The analysis of the results
shows a change of sign of MR as a function of $k_F d_1$ for $\epsilon<0.5\Delta$ which indicates a change
in the relative magnitudes of $G_P$ and $G_{AP}$. Furthermore, the small subgap values of the MR indicate
the inefficiency of the spin polarized transport operated by the Cooper pairs. On the other side, above
the superconducting gap the quasi-particles transport efficiently provides spin polarized currents and thus MR values ranging from $-10\%$ up to $20\%$ are observed.
The low values of MR below the gap are due to a thickness of the superconducting region larger than the coherence length $\xi$.
Indeed, the curves in Fig.\ref{fig:fig5} are obtained for  $d_1 \geq 3\xi$,
i.e. for a thickness such that the quasi-particles current coming from the normal leads is almost fully converted in non-polarized supercurrent.
The latter point is evident in Fig.\ref{fig:fig6} which presents the MR as a function of the size $k_F d_1$ of the superconducting region
computed for the following set of parameters:
$\epsilon/\Delta=0.01$, $z_1=z_2=z_3=0$, $\Gamma=0.5$, $\eta=1/200$, $|h_z|=0.5$, $\theta=0$.
As the superconducting spacer becomes larger than $3\xi$ (i.e. $k_Fd_1=600$) a strong suppression of the MR is observed for all the curves, while below this threshold the quasi-particles current is not efficiently converted in unpolarized supercurrent leading to a residual polarization responsible for sizeable values of MR ($\approx \pm 10\%$).
The latter results imply that a competition between superconducting and magnetic properties becomes relevant for $d_1 <2\xi$, i.e. $d_1 <40$nm for Nb superconductors. This result is consistent with that found in Ref.[\onlinecite{cuoco}] and the above conditions can be easily tackled in nanostructured devices\cite{nb_py_nanowire_exp}.\\
Regarding the oscillations observed they come from the formation of resonant states below the gap and their period is of the order of the coherence length.
From the analysis above it is evident that
the behavior of the MR is related to the amount of polarized current transmitted to the free-layer.  This quantity on turn
depends on (i) the efficiency of the fixed magnetic layer in polarizing the particle current  and
(ii) on the transmission of the polarized current produced by the polarizer (i.e. the fixed layer) through the spacer region.  Point (i) is investigated in Fig.\ref{fig:fig7} where the MR is reported as a function of the size $k_F d_1$ of the superconducting region
for different values of $|h_z|$ setting the model parameters as follows: $\epsilon/\Delta=0.01$, $z_1=z_2=z_3=0$, $\Gamma=0.5$, $\eta=1/200$, $\theta=0$, $k_F d_2=650$. Apart from the general aspect similar to the one of Fig.\ref{fig:fig6}, one observes that an
increasing of the Zeeman energy $h_z$ of the fixed layer produces higher values of MR for a superconducting spacer width smaller than $k_Fd_1=400$.
Furthermore, the behavior of the MR as a function of $h_z$ is expected to be proportional to $\Gamma h_z \cos(\theta)$, i.e. the scalar product of the magnetic momenta of the ferromagnets. This is found in Fig.\ref{fig:fig8} where we plot the MR as a function of  $|h_z|$ for the other model parameters:
$\epsilon/\Delta=0.01$, $z_1=z_2=z_3=0$, $\Gamma=0.5$, $\eta=1/200$, $\theta=0$, $k_F d_2=650$, $k_F d_1=200$.
The analysis of the figure shows a linear behavior with respect to $|h_z|$ with a slope proportional to  $\Gamma \cos(\theta)$,
while an additional oscillating pattern is observed. Such superimposed oscillations depend on the interface potentials and their amplitude increases at increasing the barrier heights $z_j$
from 0 up to 0.1.  In higher dimension (2D or 3D) we do expect that
 interface disorder can reduce the amplitude of such oscillations.
Finally the quantum size effects are displayed in Fig.\ref{fig:fig9ab} where a density-plot of the MR in the plane $(k_Fd_1,k_Fd_2)$ is shown for the model parameters:  $\epsilon/\Delta=0.01$, $z_1=z_2=z_3=0$, $\Gamma=0.5$, $\eta=1/200$, $\theta=0$ and $|h_z|=0.3$ (upper panel) or $|h_z|=0.5$ (lower panel). The dimensionless sampling step adopted in the numerical simulations is $10$ (i.e. 1 nm) for the upper panel and $25$ (i.e. 2.5 nm) for the lower panel\cite{nota3b}.
The overall behavior of the curves presented in Fig.\ref{fig:fig9ab} show oscillating patterns and a change of sign of the MR
as a function of the geometric parameters of the system. The comparison between the upper and lower panel shows the effect of the magnetic energy $|h_z|$ of the fixed layer  in rotating the wave front of the curves. This is particularly evident for the MR as a function of $k_Fd_2$ (i.e. the length of the fixed layer) for a fixed size of the superconducting spacer. This dependence can represent a relevant information for the experiments.\\

\subsection{Spin-torque}
Up to now we focussed our attention on the MR; however an additional probe of the spin polarized transport through the system is
provided by the spin torque ($T_{\perp,||}$) acting on the free-layer, see Eq.(\ref{eq:torque-free-layer}).
Despite the few experimental reports concerning the direct measurement of this observable, recently difficulties in making quantitative measurements of the spin-torque seem to be overcome. In particular magnitude and direction of the spin torque have recently been measured in magnetic tunnel junction\cite{sankey_nat_phys_torque08} leading to a substantial understanding of the angular momentum transfer in these systems.
These devices are of primary interest for the applications and
represent excellent probes of the possibility to
electrically control (using dc or ac signals) the magnetic degrees of freedom (i.e. the free layer magnetization). Within
this framework, the study of superconducting spin valves
(as the one depicted in Fig.2) can clarify the mechanism
involved in the angular momentum transfer through a
thin superconducting layer thus constituting a complementary tool in investigating the interplay between superconductivity and magnetism. A systematic analysis
of these structures, also including different symmetries
of the superconducting order parameters, could be useful
to probe exotic pairings and their ability in supporting
spin polarized current.

To start this study, we perform an analysis of the s-wave case here.
In the following we take $eV\rightarrow 1$ and thus
the quantities $T_{\perp,||}$ in units of $eV$ coincide with the derivative of the spin torque with respect to the bias in the linear response regime,
i.e. the so-called torkance.
\begin{figure}[t]
\centering
\includegraphics[scale=0.5]{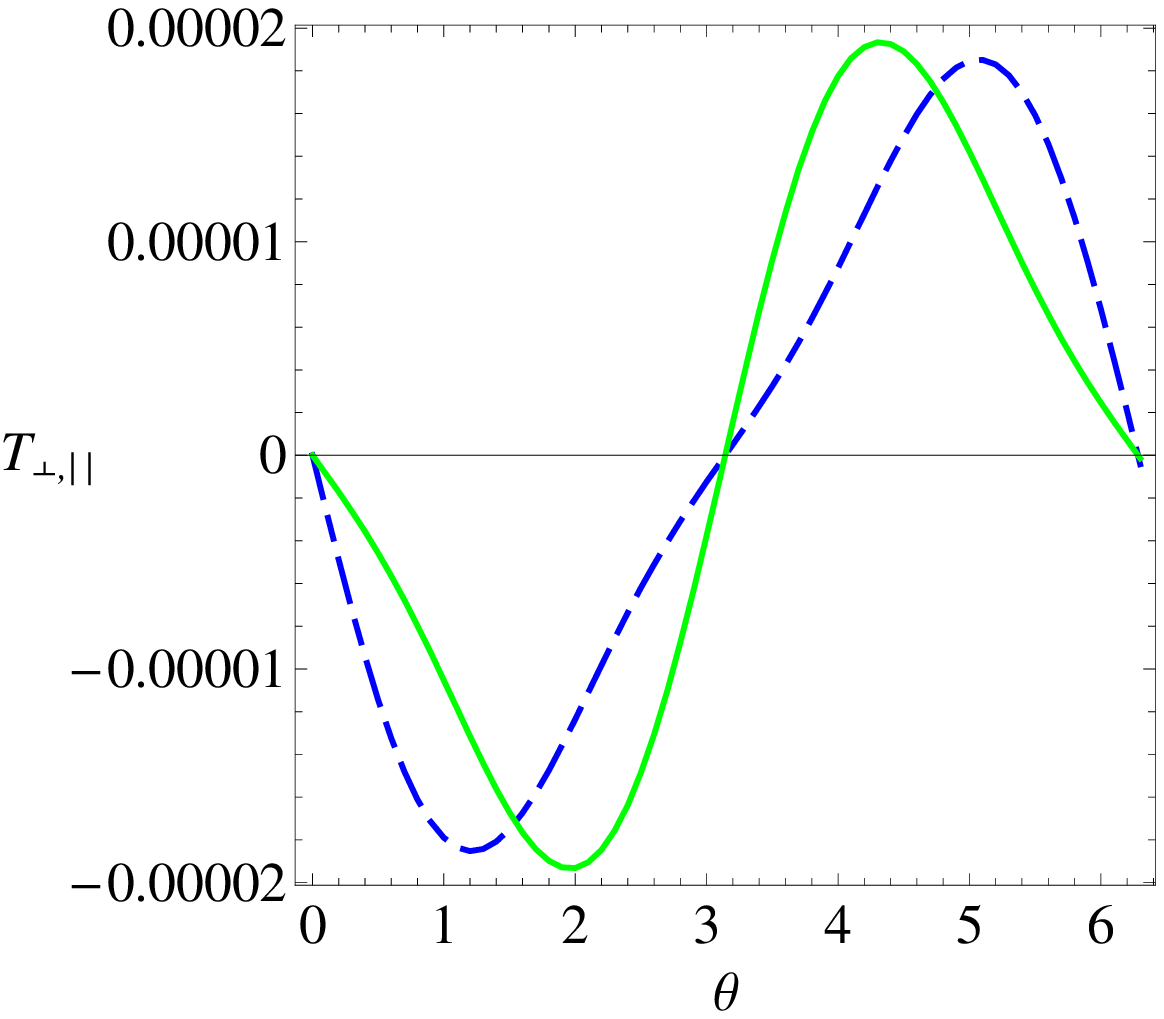}\\
\vspace{0.5cm}
\includegraphics[scale=0.5]{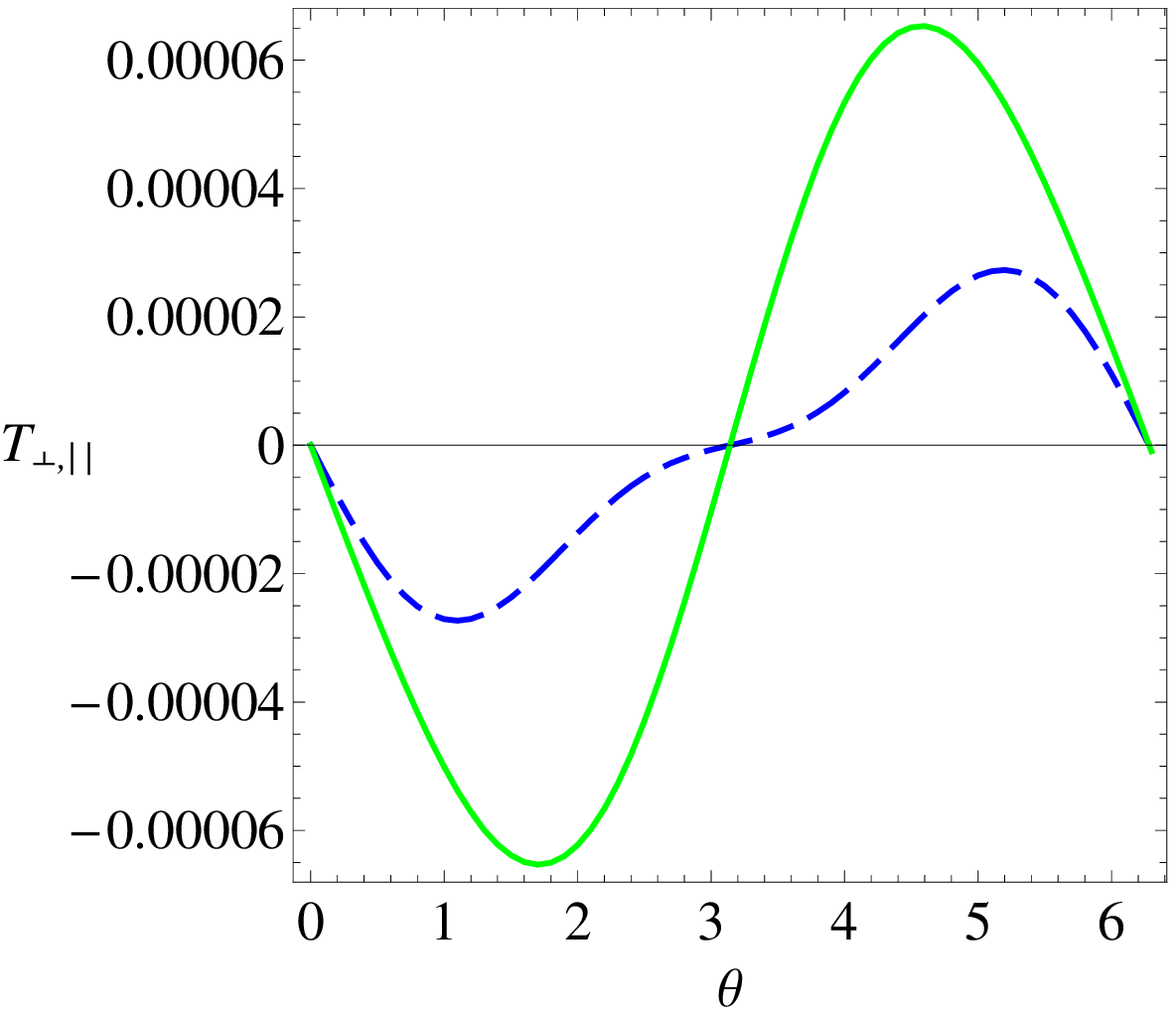}\\
\vspace{0.5cm}
\includegraphics[scale=0.5]{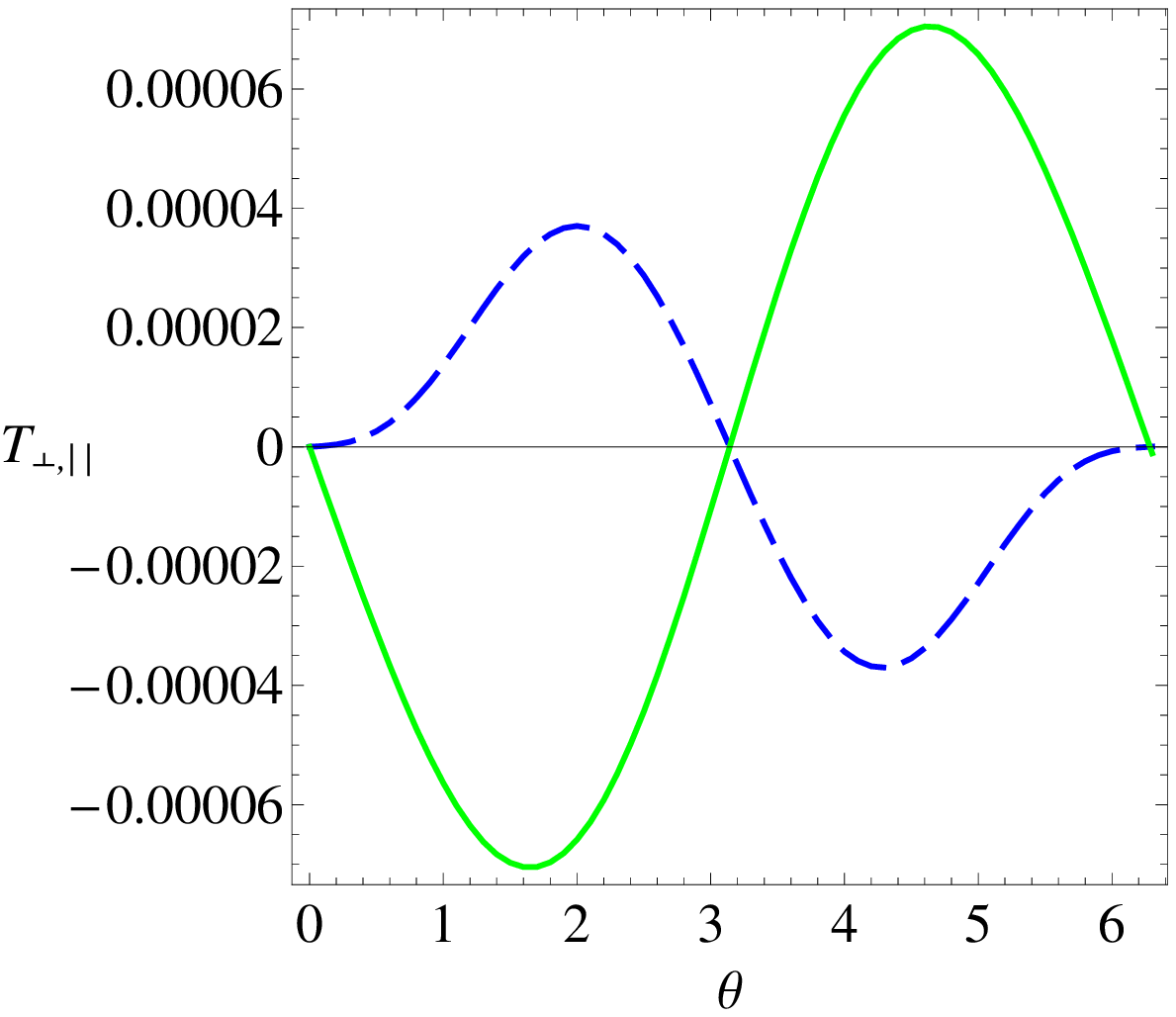}\\

\caption{Parallel (dashed line) and perpendicular (full line) component of the spin torque $T_{\perp,||}$ as a function of $\theta$ computed for the model parameters: $\epsilon/\Delta=0.01$, $h_z=0.5$, $\Gamma=0.5$, $\eta=1/200$, $\theta=\pi/2$, $k_F d_1=300$, $k_F d_2=300$, while $z=0.1$ for the upper panel, $z=0.3$ for the middle panel and $z=0.6$ for the lower panel.}
\label{fig:fig11abc}
\end{figure}
In Fig.\ref{fig:fig11abc} we report the parallel (dashed line) and perpendicular (full line) component of the spin torque $T_{\perp,||}$ as a function of $\theta$ computed setting the model parameters as follows: $\epsilon/\Delta=0.01$, $h_z=0.5$, $\Gamma=0.5$, $\eta=1/200$, $\theta=\pi/2$, $k_F d_1=300$, $k_F d_2=300$, where we use $z=0.1$ for the upper panel, $z=0.3$ for the middle panel and $z=0.6$ for the lower panel. The spin torque components present an almost sinusoidal behavior as a function of the magnetizations angle $\theta$ and thus exhibit vanishing values for $\theta=0,\pi$. The maximum values of $T_{\perp,||}$ are observed close to $\theta=\pm \pi/2$. By analyzing Fig.\ref{fig:fig11abc} we do observe an increasing of the maximum value of $T_{\perp}$ and a change of sign of $T_{||}$ going from the upper to the lower panel (i.e. by increasing $z$ from $0.1$ up to $0.6$) . The latter behavior is attributed to the difference of spin polarized currents at the interface.  The maximum (minimum) value of $T_{\perp}$ in the lower panel (see Fig.\ref{fig:fig11abc}) close to $\theta=-\pi/2$ ($\theta=\pi/2$) takes an absolute value of $0.1 \mu$eV in the presence of an applied bias of $1.5$meV. This value is of the same order of magnitude of that obtained in the case of nonsuperconducting spin-valves (see for instance Ref.[\onlinecite{kalitsov09}]). This fact points out that
nanostructured superconducting material can support a spin polarized particles transport in agreement with recent experimental findings, see e.g. Ref.[\onlinecite{nb_py_nanowire_exp}]. \\
However the values of the spin-torque strongly depend on the interface properties, i.e. on the parameters $z_j$ in our model,
 and thus a comparison with the experimental data can be done only by considering  $z_j$ as phenomenological fitting parameters.
 The behavior of the spin torque $T_{\perp,||}$ as a function of $z$ (where we set $z_1=z_2=z_3=z$) for different thickness of the SC layer is shown in Fig.\ref{fig:fig12abc}
for the choice of parameters:  $\epsilon/\Delta=0.01$, $h_z=0.5$, $\Gamma=0.5$, $\eta=1/200$, $\theta=\pi/2$, $k_F d_2=300$.
All the curves present maximum values of the torkance close to $z \sim 1$, while the maximum value of the spin torque component
$T_{\perp}$ is in the range $1-2\mu$eV. For highest values of $z$, i.e. $z>1$, $T_{\perp,||}$ start to decrease as
an effect of vanishing particles flux through the interfaces.
 On the experimental side, all our analysis can represent an efficient way of detecting the spin polarized effects in the magnetic/superconducting heterostructures,
despite the experimental difficulties of engineering reproducible interfaces.
\begin{figure}[t]
\centering
\includegraphics[scale=0.57]{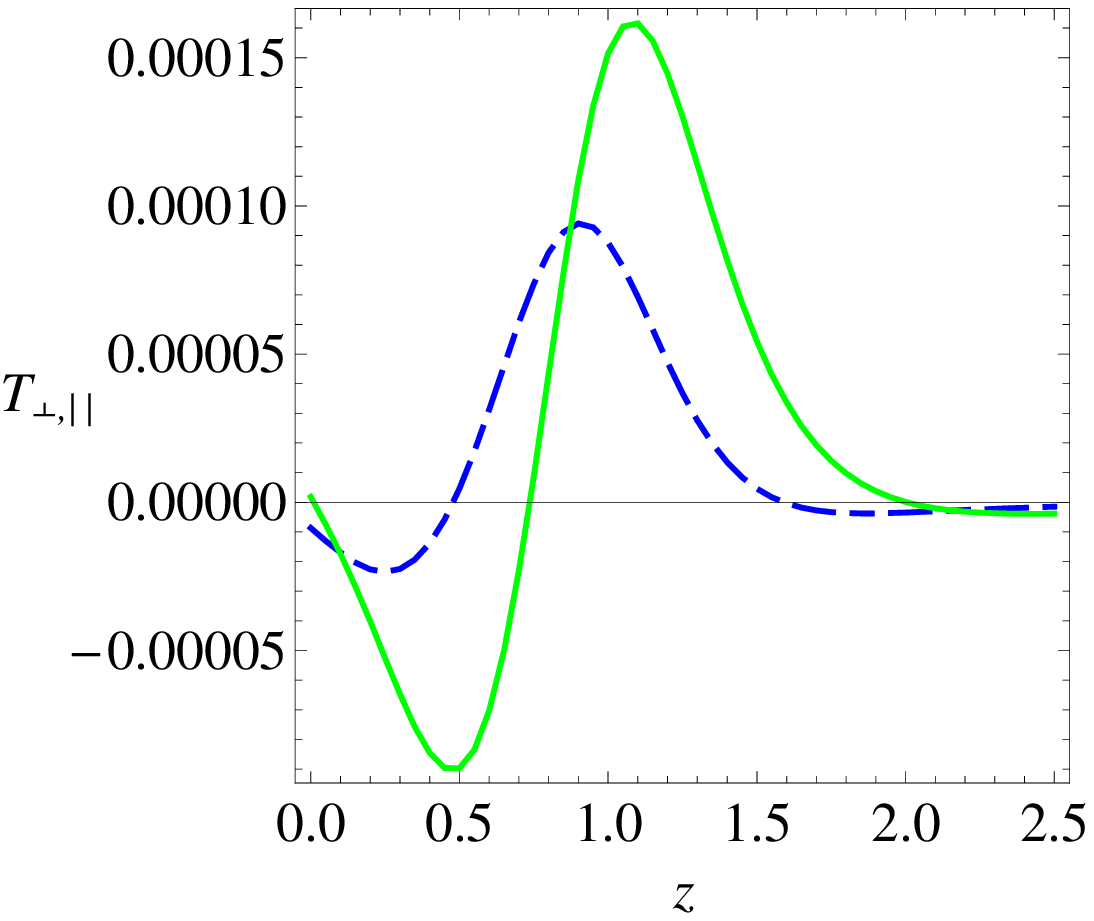}\\
\vspace{0.5cm}
\includegraphics[scale=0.55]{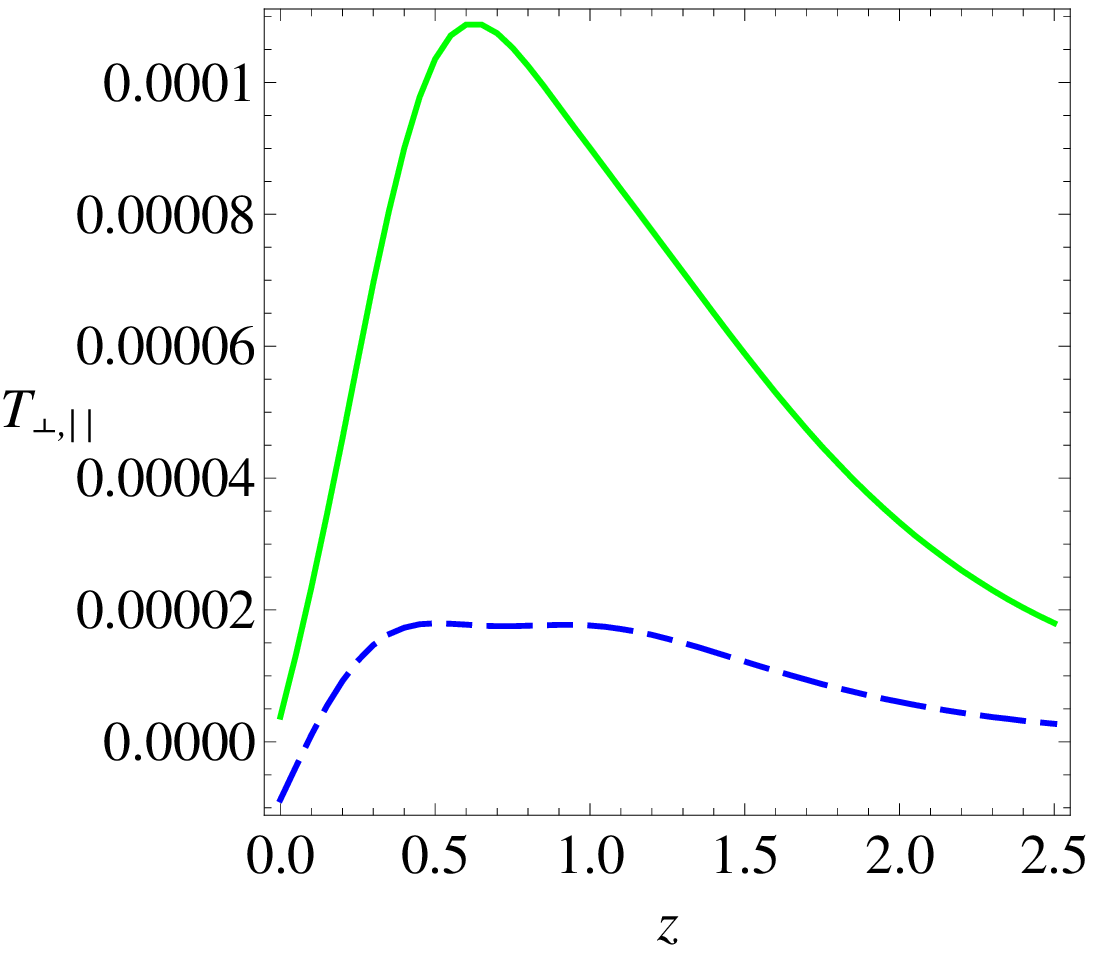}\\
\vspace{0.5cm}
\includegraphics[scale=0.55]{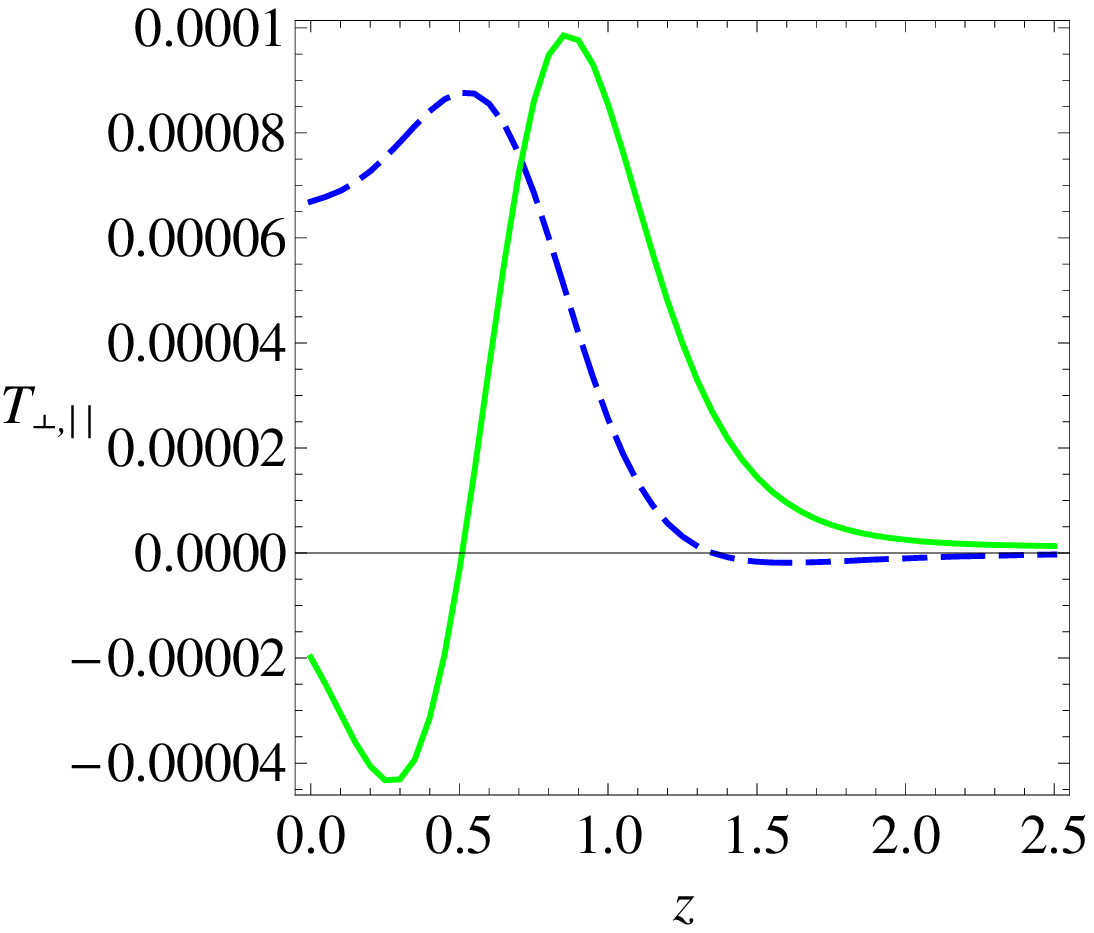}\\

\caption{Parallel (dashed line) and perpendicular (full line) component of the spin torque $T_{\perp,||}$ as a function of the interface potential $z$ ($z_1=z_2=z_3=z$) computed setting the model parameters as follows: $\epsilon/\Delta=0.01$, $h_z=0.5$, $\Gamma=0.5$, $\eta=1/200$, $\theta=\pi/2$, $k_F d_2=300$, while $k_F d_1=300$ for the upper panel, $k_F d_1=380$ for the middle panel and $k_F d_1=400$ for the lower panel.}
\label{fig:fig12abc}
\end{figure}
\section{Conclusions}
\label{sec:conclusions}
In this work a scattering field theory for quasi-one-dimensional magnetic heterostructures containing s-wave superconducting regions has been developed. The second quantized form of the scattering fields for the {\it spinful case} allows a direct link between physical observable and their relation to the scattering matrix describing the system. Our formalism take fully into account Andreev reflections in the presence of spin-flip phenomena. We formally derived spin and charge current and all the quantities related to the linear response to an applied voltage bias $V$, i.e. the conductance and the torkance. In particular, it has been pointed out that, in deriving the spin current, the charge conservation through the system must be monitored in order to guarantee the conservation laws. Indeed, the spin current is not conserved due to the presence of a spin-transfer torque acting on the local magnetization of the free-layer and thus a  violation of the charge conservation law may artificially change the spin current gradient. The above change modifies the spin torque in a quantitative way.
As for the observables, in the second part of the paper, we derived the conductance and the magnetoresistance of a superconducting spin-valve and analyzed all the relevant quantum size  and coherent effects. Our analysis showed evidence of Andreev reflections in the subgap transport at varying the Zeeman interaction revealing the importance of the spin-flip processes. As for the magnetoresistance we analyzed quantum size effects due to the superconducting layer thickness and showed that it displays a strong oscillatory and non-monotonous behavior as a function of the interlayer width. A peculiar interplay between superconducting and spin polarized transport properties becomes evident for thickness of the order of the superconducting coherence length.
As a probe of the spin-polarized transport we analyzed the spin-torque in the linear response regime and characterized its behavior as a function of the interface transparencies and direction of the magnetization between the fixed and the free layer. It has been found that the torque and magnetoresistance are both strongly enhanced by a non-zero barrier height at the interfaces.
Our analysis can
provide an efficient way of detecting spin polarized transport in experiments on magnetic/superconducting heterostructures helping some
basic understanding and stimulating further studies.

\appendix
\section{Tensor product}
\label{app:tensor-product}
In this work the sign $\otimes$ is employed to define the Kronecker product or tensor product of matrices.
Given the matrices $\mathcal{A}$ and $\mathcal{B}$ the matrix $\mathcal{C}=\mathcal{A}\otimes\mathcal{B}$ is obtained as follows:
\begin{equation}
\mathcal{C}=\left(
              \begin{array}{ccc}
                \mathcal{A}_{11} \mathcal{B}& ... & \mathcal{A}_{1n}\mathcal{B}\\
                \mathcal{A}_{m1}\mathcal{B}& ... & \mathcal{A}_{mn} \mathcal{B}\\
              \end{array}
            \right),
\end{equation}
where the size of $\mathcal{A}$ is $m \times n$. According  to the above definition, provided that $|e\rangle=(1,0)^t$ and $|-\rangle=(0,1)^t$, we get, for instance, $|e\rangle \otimes |-\rangle=(0,1,0,0)^t$.

\section{Scattering field in momentum-representation}
\label{app:scattering-field}

Within the scattering approach one assumes that far from the scattering center the particle is free and its linear momentum $p=\hbar k$ is a good quantum number for labeling the scattering states. According to this, the scattering field can be expanded in the eigenstates of the linear momentum operator $\hat{\mathcal{P}}=\sum_{\beta}P_{\beta\beta} \otimes \eta_{\beta} \mathbb{I}_{sp}(-i\hbar \partial_x)$. The eigenstates of $\hat{\mathcal{P}}$ in our tensor product notation are defined by
\begin{equation}
\hat{\mathcal{P}}| \Psi_{\beta \sigma k}(x)\rangle =\hbar k | \Psi_{\beta \sigma k}(x)\rangle,
\end{equation}
where $| \Psi_{\beta \sigma k}(x)\rangle=(\sqrt{2\pi})^{-1}|\beta\rangle \otimes |\sigma\rangle e^{i\eta_{\beta}k x}$ and $1/\sqrt{2\pi}$ is a normalization factor. These set of states satisfy the completeness relation:
\begin{equation}
\sum_{\beta\sigma}\int dk |\Psi_{\beta \sigma k}(x)\rangle \langle \Psi_{\beta \sigma k}(x')|=\mathbb{I}_{4\times 4}\delta(x-x'),
\end{equation}
where the identity operator is written as $\mathbb{I}_{4\times4}=\sum_{\beta\sigma}P_{\beta\beta}\otimes |\sigma \rangle \langle \sigma|$. The generic wave-function $|\Psi(x)\rangle=\sum_{\beta,\sigma}\phi_{\beta \sigma}(x)|\beta\rangle\otimes|\sigma\rangle$, can be written in the basis set of the eigenstates of $\hat{\mathcal{P}}$ as:
\begin{equation}
|\Psi(x)\rangle=\sum_{\beta \sigma}\int \frac{dk }{\sqrt{2\pi}}\Phi_{\beta \sigma}(k)|\beta \rangle \otimes |\sigma \rangle e^{i \eta_{\beta}k x},
\end{equation}
where the coefficients $\Phi_{\beta \sigma}(k)=\int dx' \phi_{\beta\sigma}(x') e^{-i\eta_{\beta}k x'}$, are related to the projection of $|\Psi(x)\rangle$ on the eigenvectors of $\hat{\mathcal{P}}$.

\section{Self-consistent determination of the chemical potential and the conductance tensor}
\label{app:conductance-tensor}

As described in the main text, in the non-symmetric case the chemical potential of the scattering region $\mu_s\neq (\mu_1+\mu_2)/2=\mu$ and thus a self-consistent computation of $\mu_s$ is required. Its calculation follows from the charge current conservation\cite{dong2003}, $\sum_i\bar{J}^{i}_0(V,\mu_s(V))=0$. Since in principle, such a condition implies the solution of an integral equation, a great simplification follows in the linear response regime in the applied voltage bias $V$. In this case the charge current flowing through the $i$-th lead is obtained as $I_i=\sum_j g_{ij} (\mu_j-\mu_s)$, where $g_{ij}$ is the conductance tensor and the charge conservation implies $\sum_i I_i=0$. Solving the latter equation (Kirchhoff's law) with respect to $\mu_s$ we have:
\begin{equation}
\label{eq:chem-pot-sc-region}
\mu_s=\frac{\sum_{ij}g_{ij}\mu_j}{\sum_{ij}g_{ij}}.
\end{equation}
From the equation above it immediately follows that in the case of a two-terminal\cite{nota4} symmetric system (where $g_{11}=g_{22}$) the chemical potential $\mu_s$ is bias independent, $\mu_s=\mu$. More generically, one can analyze the potential drop to the left and right junction, i.e.:
\begin{equation}
\mu_j-\mu_s=q_e V \lambda_j,
\end{equation}
where the coefficients $\lambda_j$  are function of $g_{ij}$ as shown below:
\begin{eqnarray}
&&\lambda_1=-\frac{g_{12}+g_{22}}{\sum_{ij}g_{ij}}\\\nonumber
&& \lambda_2=\frac{g_{21}+g_{11}}{\sum_{ij}g_{ij}}.
\end{eqnarray}
Observing that $\lambda_2-\lambda_1=1$ one correctly recovers that $\mu_2-\mu_1=q_e V (\lambda_2-\lambda_1)=q_e V$, while for a symmetric system  $\lambda_j=(-)^j/2$.
From the definitions above one immediately infers that the electrochemical potential of the scattering region $\mu_s$ is displaced from $\mu=(\mu_1+\mu_2)/2$ according to the relation:
\begin{equation}
\label{eq:mu-s}
\mu_s=\mu+\frac{q_e V}{2}\Bigl[\frac{g_{22}-g_{11}}{\sum_{ij}g_{ij}}\Bigl].
\end{equation}
Noticing that in the linear response regime $I_i=G_iV$ where $G_i=\sum_j g_{ij}\lambda_j$ and using the expression above for $\lambda_j$ one obtains\cite{nota5}:
\begin{equation}
\label{eq:two-probe cond}
G=\frac{g_{22}g_{11}-g_{21}g_{12}}{\sum_{ij}g_{ij}}.
\end{equation}
For symmetric systems the relation above can be simplified as
\begin{equation}
\label{eq:g_sym_bis}
G_{sym}=(g_{11}-g_{12})/2
\end{equation}
Let us note that Eq.(\ref{eq:g_sym_bis}) and Eq.(\ref{eq:g_sym}) in the main text reproduces the result given in Eq.(11) of Ref.[\onlinecite{dong2004}] using the Lambert's method. However, since we are considering a one-dimensional structure in place of the bidimensional one, the angular integration $\int d\theta \cos(\theta)[\cdot \cdot \cdot ]$ is not present in our result.

\section{Boundary conditions of the scattering problem}
\label{app:boundary}
To determine the scattering matrix coefficients one has to use the mode-matching technique as formulated in the theory of quantum wave-guides. According to this method, the BdG equation  is solved in each branch and the resulting eigenmodes are used to expand the scattering wave-function. Each wave-function is than determined by imposing proper boundary conditions\cite{boundary-conditions}.
E.g. in the presence of a single particle magnetic potential $U(x)=\gamma \delta(x) \hat{n} \cdot \vec{\sigma}$, $\hat{n}=(n_x,n_y,n_z)$ being the unit vector describing the magnetization direction ($|\hat{n}|^2=1$), the BdG wavefunction $\Psi(x)=(u_{\uparrow}(x),u_{\downarrow}(x),v_{\uparrow}(x),v_{\downarrow}(x))^t$ must satisfy the following boundary conditions:
\begin{eqnarray}
&&\Psi(x=0^+)=\Psi(x=0^-)\\\nonumber
&&\partial_x \Psi(x=0^+)-\partial_x \Psi(x=0^-)=\frac{2m \gamma}{\hbar^2}\mathcal{A}\Psi(x=0^+),
\end{eqnarray}
where the $4 \times 4$ matrix $\mathcal{A}$ is defined as follows:
\begin{equation}
\mathcal{A}=\left(
              \begin{array}{cc}
                \hat{n} \cdot \vec{\sigma} & 0 \\
                0 & \hat{n} \cdot \vec{\sigma}^{\ast} \\
              \end{array}
            \right).
\end{equation}
In the case of a non-magnetic potential $U(x)=\gamma\delta(x)\mathbb{I}_{sp}$ the previous boundary conditions must be modified substituting $\mathcal{A}$ with the $4 \times 4$ identity $\mathbb{I}_{4 \times 4}$, i.e. $\mathcal{A} \rightarrow \mathbb{I}_{4 \times 4}$. In the absence of potential, i.e. $\gamma=0$, the boundary conditions imply the continuity of the BdG wavefunction and its derivative.

\section*{Acknowledgements}
The authors wish to thank G. Annunziata, C. Attanasio, M. Cuoco, A. Di Bartolomeo, F. Giubileo , G. Lambiase and A. Sorgente
for helpful discussions during the preparation of the present work.
\bibliographystyle{prsty}

\begin{thebibliography}{99}
\bibitem{nota1} Notice that the spinorial particle-hole notation adopted in this work follows from the Bogoliubov field transformation $\psi_{\sigma}(x)=\sum_n [u_{n\sigma}(x) \gamma_n+v_{n\sigma}^{\ast}(x)\gamma_n^{\dagger}]$. Using the Bogolons rapresentation the BCS Hamiltonian can be diagonalized assuming the form $H_{BCS}=\sum_n E_n \gamma^{\dagger}_n\gamma_n+E_g$. On this point see also Ref.[\onlinecite{degennes_book}].
\bibitem{nota2}Notice that a more precise notation should be of the form $|\beta\rangle\otimes|\sigma\rangle_{\beta}$, where $|\sigma\rangle_{h}=|\sigma\rangle_{e}^{\ast}$ provided that $|\sigma\rangle_{e}$ is eigenvector of the spin dependent part of the single particle potential (proportional to $\hat{n}\cdot \vec{\sigma}$). However, in case of non-magnetic leads the scattering events can be described using the eigenstates of $\sigma_z$ and thus the notation can be simplified as done in the main text.
\bibitem{nota3} In general the scattering relation takes the form $b^{\sigma}_{i\beta}(t)=\sum_{i'\sigma'\beta'}\int dt'S^{\beta\beta'}_{ii'\sigma\sigma'} (t,t') a^{\sigma'}_{i'\beta'}(t')$. Performing an istantaneous scattering approximation, i.e. $S(t,t')\approx \delta(t-t')S(t)$, a semplified scattering relation is obtained. Within this framework the static and adiabatic scatterer regime can be described.
\bibitem{nota3b} Despite the small step considered, one should keep in mind that this sampling could not be totally suited to discern oscillations with period less that the sampling one and then reveal the true multiple frequencies contained in the curve.
\bibitem{nota4}Notice that for a two-terminal system the Onsager's symmetry $g_{12}=g_{21}$ is respected.
\end{thebibliography}

\end{document}